\documentclass[%
reprint,
showkeys,
nofootinbib,
amsmath,amssymb,
aps,
prl,
]{revtex4-1}
\usepackage[marginal]{footmisc}
\usepackage{graphicx}
\usepackage{dcolumn}
\usepackage{bm}
\usepackage[colorlinks,allcolors=blue,CJKbookmarks=True]{hyperref}

\begin{document}

\title{ Large anisotropies of the stochastic gravitational wave background from cosmic domain walls}

\author{Jing Liu$^{1,2}$}
\email{liujing@itp.ac.cn}

\author{Rong-Gen Cai$^{3,2,1}$}
\email{cairg@itp.ac.cn}

\author{Zong-Kuan Guo$^{3,2,1}$}
\email{guozk@itp.ac.cn}

\affiliation{$^{1}$School of Fundamental Physics and Mathematical Sciences, Hangzhou Institute for Advanced Study, University of Chinese Academy of Sciences, Hangzhou 310024, China
}

\affiliation{$^{2}$School of Physical Sciences, University of Chinese Academy of Sciences,
	No.19A Yuquan Road, Beijing 100049, China}

\affiliation{$^{3}$CAS Key Laboratory of Theoretical Physics, Institute of Theoretical Physics,
 Chinese Academy of Sciences, P.O. Box 2735, Beijing 100190, China}

\begin{abstract}
We investigate the stochastic gravitational wave background~(SGWB) from cosmic domain walls~(DWs) caused by quantum fluctuations of a light scalar field $\phi$ during inflation. Perturbations of $\phi$ remain almost constant after leaving the Hubble horizon. The probabilities of the two domains depend on the averaged value of $\phi$ at each large scale region, leading to large scale perturbations of DW energy density and large anisotropies in the SGWB. We find that the angular power spectrum is scale-invariant and at least of the order of $10^{-2}$, which is expected to be detected by future gravitational-wave~(GW) detectors. Since we have not detected primordial GWs yet, anisotropies of the SGWB could help us to verify the rationality of inflation and determine the energy scale of inflation. 
\end{abstract}

\maketitle

\emph{Introduction}. 
The direct detection of GWs produced by black hole binary merger events opens a new era of GW astronomy and cosmology~\cite{Cai:2017cbj}. GW sources may also exist in the early universe, such as phase transition~\cite{Steinhardt:1981ct,Kosowsky:1991ua,Cai:2017tmh}, preheating~\cite{Khlebnikov:1997di,Liu:2017hua}, topological defects~\cite{Vachaspati:1984gt,Auclair:2019wcv,Saikawa:2017hiv,Zhou:2020ojf}, large amplitude scalar perturbations~\cite{Ananda:2006af,Baumann:2007zm,Kohri:2018awv}. Large density perturbations inside the Hubble horizon during these physical process constitute sources producing substantial GWs. Due to the weakness of gravitational interaction, GWs produced by those sources can be directly observed by the GW detectors, and then provide essential clues about the history of the early universe and the high energy physics beyond the Standard Model of particle physics.


The Hubble horizon extends many orders of magnitude after the GW production, so GWs from the early universe are superposed from a great many individual sources, forming SGWBs with stochastic propagation directions, similar to the cosmic microwave background~(CMB). Recently, the NANOGrav collaboration claimed a strong evidence of a stochastic common-spectrum process~\cite{Arzoumanian:2020vkk}, which could be interpreted as a SGWB.
The energy spectrum of SGWBs is not a blackbody spectrum, and the profiles of the GW energy spectrums receive extensive investigations as characteristics of different sources~\cite{Kuroyanagi:2018csn}.
Anisotropies of SGWBs receive more attention recently, both from the sources~\cite{Bethke:2013aba,Bethke:2013vca,Geller:2018mwu,Jenkins:2018lvb}, during propagation~\cite{Contaldi:2016koz,Bartolo:2019oiq,Bartolo:2019yeu,DallArmi:2020dar}, most of which are still challenging to observe~\cite{Allen:1996gp,Seto:2004np,Hotinli:2019tpc,Chu:2020qiw,Mentasti:2020yyd}. Different from $10^{-18}\mathrm{Hz}$ primordial GWs which result in the CMB B-mode polarizations, the peak frequencies of those SGWBs could be within the sensitivity bands of various detectors such as aLIGO~\cite{TheLIGOScientific:2014jea}, LISA~\cite{Audley:2017drz}, Taiji~\cite{Guo:2018npi} and SKA~\cite{Carilli:2004nx}, meanwhile, GW energy density perturbations is at the CMB scales. Since the inflationary energy scale is too low for the Planck satellite to find primordial GWs~\cite{Liddle:1993ch,Guo:2010mm,Akrami:2018odb}, anisotropies of SGWBs can also be used as a probe of inflation. 

During inflation there may exist an extra light scalar field, for example the Higgs field, the string axions. After inflation the light field rolls down the potential when the Hubble parameter becomes smaller than the effective mass of the field, and accounts for both curvature perturbations and entropy perturbations~\cite{Enqvist:2001zp,Lyth:2002my,Moroi:2001ct}.
One interesting case is the light field also produces GWs or affects the GW production process.
Quantum fluctuations of the light field in the early times of inflation can lead to large scale perturbations of the energy density of the GW sources, generating anisotropic SGWBs. Since quantum fluctuations of the extra light field are not constrained by the $10^{-5}$ CMB temperature anisotropies, we have the opportunity to observe large scale SGWB anisotropies by GW observers in plan.
In this letter, we focus on the anisotropic SGWB from cosmic DWs, where the discrete symmetry of the effective potential is spontaneous broken. At the beginning of inflation, the initial value of the light scalar $\phi$ is arbitrary, and in some regions $\phi$ crosses the potential barrier by quantum fluctuations. As the Hubble parameter decreases after inflation, $\phi$ finally rolls down the potential and the DWs form. Different from the previous works, the probabilities of the two domains are unequal, since the initial value is biased and the Hubble parameter during inflation is not much larger than the vacuum expectation value of $\phi$. Perturbations of $\phi$ which leave the Hubble horizon in the first several efoldings cause that the averaged values in large scale regions deviate from the initial value of $\phi$, which result in large scale perturbations of the probability ratio of the two domains, the DW energy density, and the GW energy density. 
To prevent DWs from overclosing the universe, we apply the models where one of the vacua is slightly lifted~\cite{Vilenkin:1981zs,Gelmini:1988sf}, motivated by the Higgs field models~\cite{Kitajima:2015nla,Krajewski:2016vbr}, axion models~\cite{Daido:2015gqa,Higaki:2016yqk} and supersymmetric models~\cite{Takahashi:2008mu,Dine:2010eb}. Depending on the annihilation time of DWs, the peak frequencies of the SGWB could lie within the sensitivity bands of the GW detectors. In some parameter space, the SGWB could explain the common-spectrum process observed by NANOGrav.
We find strong anisotropies of at least $\sim0.1$ variations at the CMB scales are predicted by our models as long as the scalar field is light during inflation, which could be used as a noval method to probe inflation.
We set $c =8\pi G=1$ throughout this letter.

\begin{figure}[h]
	\includegraphics[width=3.2in]{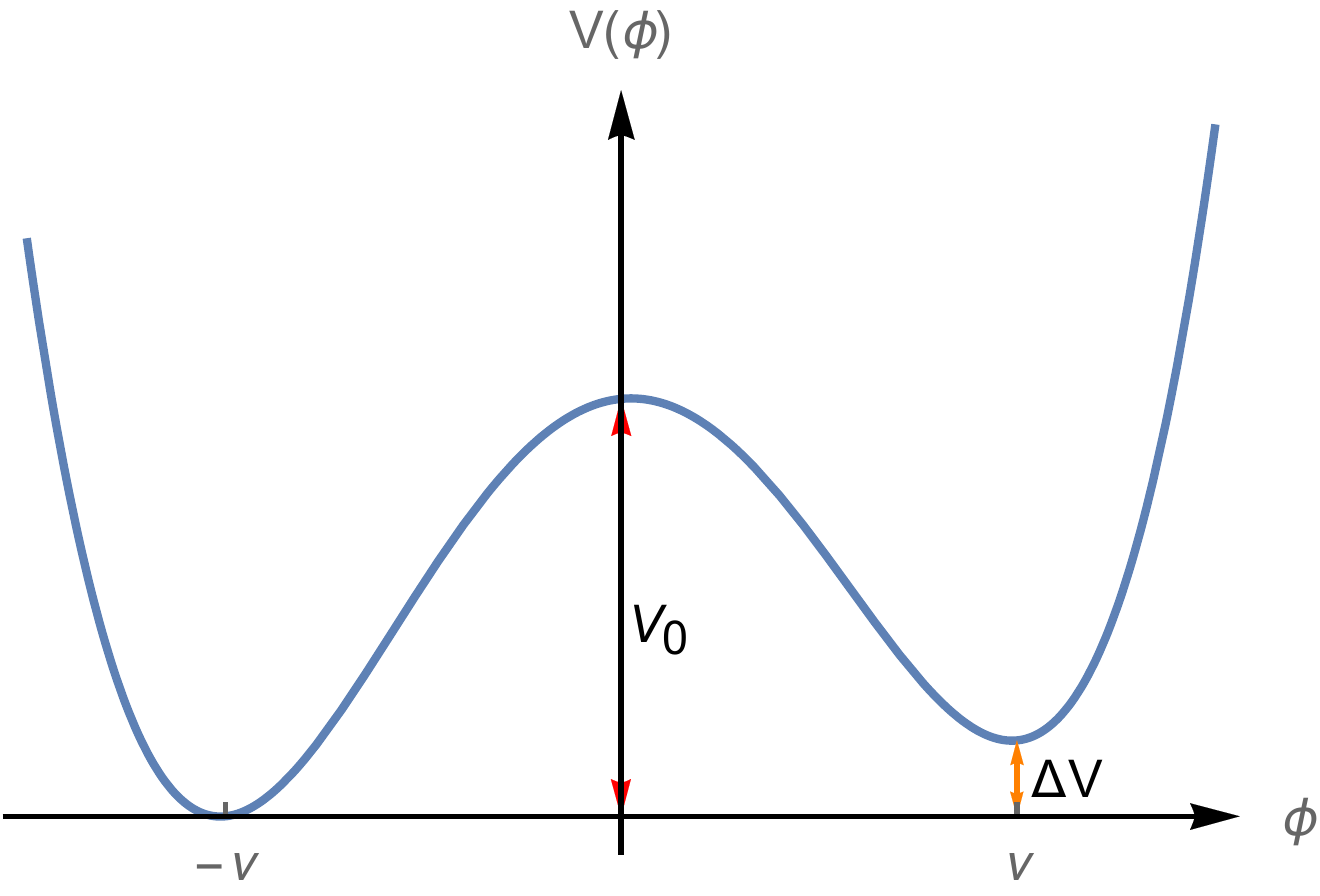}
	\caption{The discrete symmetry breaking effective potential with a biased term.}
	\label{fig:Vphi}
\end{figure}

\emph{GWs from cosmic DWs}.
Consider a scalar field $\phi$ with an effective potential $V(\phi)$ in Einstein gravity,
where the minima of $V(\phi)$, $\phi=\pm \nu$, are separated by the potential barrier $V_{0}$ as depicted in Fig.~\ref{fig:Vphi}. Let $\phi(z)$ denote the static planar DW solution perpendicular to the $z$-axis in Minkowski space. The tension of DWs, $\sigma$, is obtained by integrating the energy density $(d\phi/dz)^{2}/2+V(\phi)$ along the direction perpendicular to the wall,
which is also the surface energy density of DWs.

To prevent DWs from overclosing the universe, one can lift one of the degenerate vacuum by $\Delta V$, so that DWs annihilate at time $t_{\mathrm{ann}}\sim \mathcal{A}\sigma/\Delta V$~\cite{Saikawa:2017hiv},
where $\mathcal{A}\approx 0.8$~\cite{Hiramatsu:2013qaa} is fixed by numerical simulation.

In the previous works, the initial value of $\phi(\mathbf{x})$ is randomly set and the probabilities of the two domains are equal. In this case, the numerical and analytical results of GWs from DWs at the radiation-dominated era is obtained in Refs.~\cite{Hiramatsu:2013qaa,Saikawa:2017hiv}. 
The energy spectrum of GWs, $\Omega_{\mathrm{GW}}(k)$, is proportional to $t^{2}$, so most energy in GWs is produced nearly at $t_{\mathrm{ann}}$. 
Then $\Omega_{\mathrm{GW}}$ at $t_{\mathrm{ann}}$ reads
\begin{equation}
\Omega_{\mathrm{GW},\mathrm{peak}}\left(t_{\mathrm{ann}}\right)=\frac{ \tilde{\epsilon}_{\mathrm{GW}} \mathcal{A}^{2} \sigma^{2}}{24\pi H^{2}\left(t_{\mathrm{ann}}\right)}\,,
\end{equation}
where $\tilde{\epsilon}_{\mathrm{GW}}\approx 0.7$ is a constant given by numerical results~\cite{Hiramatsu:2013qaa}. According to Ref.~\cite{Hiramatsu:2013qaa}, $\Omega_{\mathrm{GW}}(k)$ scales as $k^{3}$ for $k<k_{\mathrm{peak}}$ and $k^{-1}$ for $k>k_{\mathrm{peak}}$.

The peak wavelength is close to the Hubble horizon size at $t_{\mathrm{ann}}$, and then redshifted by the expansion of the universe, so the peak frequency $f_{\mathrm{peak}}$ and the peak amplitude at the present time $t_{0}$ reads
\begin{equation}
\begin{split}
f_{\mathrm{peak}}=\left(\dfrac{H^{2}(t_{\mathrm{ann}})}{H_{0}^{2}\Omega_{\mathrm{rad}}(t_{0})}*\left(\dfrac{g_{*\mathrm{ann}}}{g_{*0}}\right)^{1/3}\right)^{-1/4}H(t_{\mathrm{ann}})\,,\\
\Omega_{\mathrm{GW,peak}}(t_{0})h^{2}=\Omega_{\mathrm{rad}}(t_{0})h^{2}\left(\dfrac{g_{*0}}{g_{*\mathrm{ann}}}\right)^{1/3}\Omega_{\mathrm{GW,peak}}\left(t_{\mathrm{ann}}\right),
\end{split}
\end{equation}
where $g_{*0}$ and $g_{*\mathrm{ann}}$ are the effective relativistic degrees of freedom at $t_{0}$ and $t_{\mathrm{ann}}$, respectively, $\Omega_{\mathrm{rad}}h^{2}=4.2\times 10^{-5}$ is the density fraction of radiation at $t_{0}$, and $H_{0}$ is the Hubble constant, $H_{0}=67.7\, \mathrm{km}\,\mathrm{s}^{-1}\,\mathrm{Mpc}^{-1}$ from the results of Planck 2018~\cite{Aghanim:2018eyx}.


\emph{Anisotropies}.
Let us then focus on how anisotropies of the SGWB are generated. Let $\phi$ denote the light field during inflation instead of the inflaton. 
Quantum fluctuations of $\phi$ during inflation lead to perturbations $\delta\phi(\mathbf{x})$ at super-horizon scales. The initial background value of $\phi$  is set to be $\phi_{i}$, while the initial cosmic time $t_{i}$ is defined by $a(t_{0})H_{0}=a(t_{i})H_{\mathrm{inf}}$, where $a(t_{0})=1$ and $H_{\mathrm{inf}}$ is the Hubble parameter during inflation. Let $P(\tilde{\phi},t)$ denotes the probability of $\tilde{\phi}$ at $t$, where $\tilde{\phi}$ is the value of spacial averaged $\phi$ inside one Hubble horizon. For the light field during inflation, $P(\tilde{\phi},t)$ reads
\begin{equation}
P(\tilde{\phi}, t)=\sqrt{\frac{2 \pi}{H^{3} (t-t_{i})}} \exp \left(-\frac{2 \pi^{2}}{H^{3} (t-t_{i})} \tilde{\phi}^{2}\right)\,,
\end{equation}
which is the solution of the Fokker-Planck equation~\cite{Espinosa:2015qea}.
During inflation $H_{\mathrm{inf}}$ is almost a constant, so the efolding numbers $N(t)\equiv \ln(a(t)/a(t_{i}))$ can be expressed as a function of $t$ as $N(t)=H_{\mathrm{inf}}(t-t_{i})$. Without loss of generality, setting $\phi_{i}>0$, the probability of $\tilde{\phi}(t)<0$ at $t$ is
\begin{equation}
\label{eq:prob}
P\left(\tilde{\phi}<0, t\right) 
=\dfrac{1}{2}\operatorname{erfc}\left(\frac{\sqrt{2} \pi \phi_{i}}{H_{\mathrm{inf}} \sqrt{N(t)}}\right)\,.
\end{equation}
The regions where $\tilde{\phi}<0$ and $\tilde{\phi}<0$ fall in different vacua when the DWs form unless $H_{\mathrm{inf}}\gg\nu$, so Eq.~\ref{eq:prob} implies that the possibilities of the two domains are different. We define $\alpha(t)\equiv \frac{\sqrt{2} \pi \phi_{i}}{H_{\mathrm{inf}}\sqrt{N(t)}}$, which is of the order of $0.1$. Large scale perturbations of $\phi$ result in large scale perturbations of $P(\tilde{\phi}<0, t)$\footnote{Since $\delta\phi(\mathbf{x})$ is caused by quantm fluctuations in the first several efoldings, $N(t)$ overestimates the efolding numbers. This deviation is negligible if we focus on anisotropies at large scales.}, 
\begin{equation}
\label{eq:deltaP}
\begin{split}
\delta P\left(\tilde{\phi}<0, t,\mathbf{x}\right)=&\dfrac{1}{2}\operatorname{erfc}\left[\alpha(t)(1+\delta\phi(\mathbf{x})/\phi_{i})\right]\\
&-\dfrac{1}{2}\operatorname{erfc}\left[\alpha(t)\right]\,.
\end{split}
\end{equation}
Here we emphesize that around $\mathbf{x}$ we apply spacial average at a scale much larger than the wavelength of GWs but much smaller than the Hubble horizon scale at $t_{0}$. Note that no matter the discrete symmetry breaking happens before or after inflation, Eq.~\eqref{eq:prob} is valid as long as $\frac{dV/d\phi}{3H}$ is negligible. If the effective mass of $\phi$ is larger than $H$ due to the thermal correction term after inflation, the background value of $\phi$ starts to oscillate around the minimum of the potential with the oscillation amplitude decreasing. We find that the probability ratio of the two domains is still fixed under this situation. Thermal fluctuations are inside the Hubble horizon before DW formation and do not affect our results at large scales.


Since the averaged radius of DWs at $t$ is comparable to the Hubble horizon size at $t$, the area of DWs is a constant for each Hubble volume where the averaged value of $\phi$ is negative.
Therefore, the energy density of DWs, $\rho_{\mathrm{DW}}$, is proportional to its area in unit volume, $\rho_{\mathrm{DW}}(t)\propto P(\tilde{\phi}<0, t)$.
Since most of the GW energy is produced nearly at $t_{\mathrm{ann}}$, $\Omega_{\mathrm{GW}}(k)$ is proportional to the energy density of the source $\rho_{\mathrm{DW}}$ at $t_{\mathrm{ann}}$,
\begin{equation}
\label{eq:OmegaP}
\Omega_{\mathrm{GW},P}(k)=2P(\tilde{\phi}<0,t_{\mathrm{ann}})\Omega_{\mathrm{GW}}(k)\,.
\end{equation}
If the possibilities of the two domains $\phi<0$ and $\phi>0$ are equal, then $P(\tilde{\phi}<0,t_{\mathrm{ann}})=1/2$ and $\Omega_{\mathrm{GW},P}(k)=\Omega_{\mathrm{GW}}(k)$.

The averaged comoving radius of DWs at $t_{\mathrm{ann}}$ is close to $k_{\mathrm{peak}}$, which implies $N(t)$ in Eq.~\eqref{eq:prob} and Eq~.\eqref{eq:deltaP} can be determined by $k_{\mathrm{peak}}$.
The peak mode leaves the Hubble horizon when $k_{\mathrm{peak}}=a(t)H_{\mathrm{inf}}$, and the Hubble constant satisfies $H_{0}= a(t_{i})H_{\mathrm{inf}}$, so the efolding number for the peak is
\begin{equation}
N_{\mathrm{peak}}=\ln(k_{\mathrm{peak}}/H_{0})\,,
\end{equation}
and $\alpha_{\mathrm{p}}\equiv\frac{\sqrt{2} \pi \phi_{i}}{H_{\mathrm{inf}}\sqrt{N_{\mathrm{peak}}}}$.

Taking into consideration the inability of the GW detectors to probe high angular resolution of the SGWB, we focus on large scale anisotropies of the SGWB. Since for each $k$-mode anisotropies of $\Omega_{\mathrm{GW}}(k)$ are proportional to perturbations of $\rho_{\mathrm{DW}}$ at $t_{\mathrm{ann}}$, anisotropies are independent of $k$ and we omit the variable $k$ in $\Omega_{\mathrm{GW}}(k)$ in the following.
Perturbations of $\Omega_{\mathrm{GW},P}$ is defined by $\delta\Omega_{\mathrm{GW},P}(\mathbf{x})\equiv(\Omega_{\mathrm{GW},P}(\mathbf{x})-\overline{\Omega_{\mathrm{GW},P}})/\overline{\Omega_{\mathrm{GW},P}}$, where the overline denotes the spacial average in total space. 
$\delta\Omega_{\mathrm{GW},P}(\mathbf{x})$ is proportional to large scale perturbations of $\phi$ in the first order
\begin{equation}
\label{eq:deltaGW}
\delta\Omega_{\mathrm{GW},P}(\mathbf{x})
=c_{1}\delta\phi(\mathbf{x})\,.
\end{equation}
Here coefficient $c_{1}$ is given by Eq.~\eqref{eq:deltaP} and Eq.~\eqref{eq:OmegaP},
\begin{equation}
\label{eq:c1}
c_{1}=\dfrac{2}{\sqrt{\pi}\phi_{i}}\dfrac{\exp(-\alpha_{\mathrm{p}}^{2})\alpha_{\mathrm{p}}}{\dfrac{1}{2}\operatorname{erfc}\left(\alpha_{\mathrm{p}}\right)}\,.
\end{equation}
Using the approximation $\mathrm{erfc}(x)\approx \frac{2}{\sqrt{\pi}}e^{-x^{2}}$ for $x\gg 1$, we have $c_{1}\approx 2\alpha_{\mathrm{p}}^{2}/\phi_{i}$ for $\alpha_{\mathrm{p}}\gg 1$.

Analogous to the Sachs-Wolfe plateau for temperature
fluctuations for small multiple $l$~\cite{Liddle:2000cg}, the angular power spectrum can be expressed in terms of the power spectrum of $\delta \Omega_{\mathrm{GW}}(\mathbf{x})$ by
\begin{equation}
\label{eq:plat}
l(l+1)C_{l}=\frac{\pi}{2}\mathcal{P}_{\mathrm{GW}}\,,
\end{equation}
where $\mathcal{P}_{\mathrm{GW}}\equiv\langle\delta\Omega_{\mathrm{GW}}^{2}\rangle$.
Since the angular power spectrum is frequency independent, combining Eq.~\eqref{eq:deltaGW}, Eq.~\eqref{eq:c1}, Eq.~\eqref{eq:plat} and the relation $\langle\delta\phi^{2}\rangle=\frac{H_{\mathrm{inf}}^{2}}{4\pi^{2}}$, we obtain
\begin{equation}
\label{eq:cl}
l(l+1) C_{l}\approx\left\{
\begin{aligned}
&\frac{\pi}{N_{\mathrm{peak}}}\alpha_{\mathrm{p}}^{2},&\alpha_{\mathrm{p}}\gg 1,\\
&\dfrac{1}{N_{\mathrm{peak}}},&\alpha_{\mathrm{p}}\ll 1. \\
\end{aligned}
\right.
\end{equation}
$N_{\mathrm{peak}}$ must be smaller than $60$, so $l(l+1)C_{l}$ is larger than $10^{-2}$.
Setting $\phi_{i}$ near the true vacuum of the potential, $\phi_{i}$ and $\nu$ are of the same order, so the inflationary scale can also be derived using Eq.~\eqref{eq:cl}, if $\nu$ is determined by the particle physics models. 

\emph{Examples}.
1) Consider the formation of DWs in the spontaneous breaking of discrete $R$ symmetries discussed in Ref.~\cite{Takahashi:2008mu}.
DWs form after the gauge interaction becomes strong at the scale $\Lambda_{c}$, the tension of DWs is $\sigma \sim \Lambda_{c}^{3}$.
The bias term $\Delta V$ is relative to the mass of gravitinos $\Delta V \sim m_{3 / 2} \Lambda_{c}^{3}$. Choosing the parameters as $\Lambda_{c}=5\times10^{10}\mathrm{GeV}$, $\phi_{i}=2H_{\mathrm{inf}}$ and $m_{3 / 2}=1\mathrm{MeV}$, the GW energy spectrum peaks at $f=1\mathrm{Hz}$ with the peak value $\Omega_{\mathrm{GW},P}(t_{0})h^{2}=\times10^{-11}$, and the angular power spectrum $l(l+1)C_{l}=0.12$.

\begin{figure}[h]
	\includegraphics[width=3.2in]{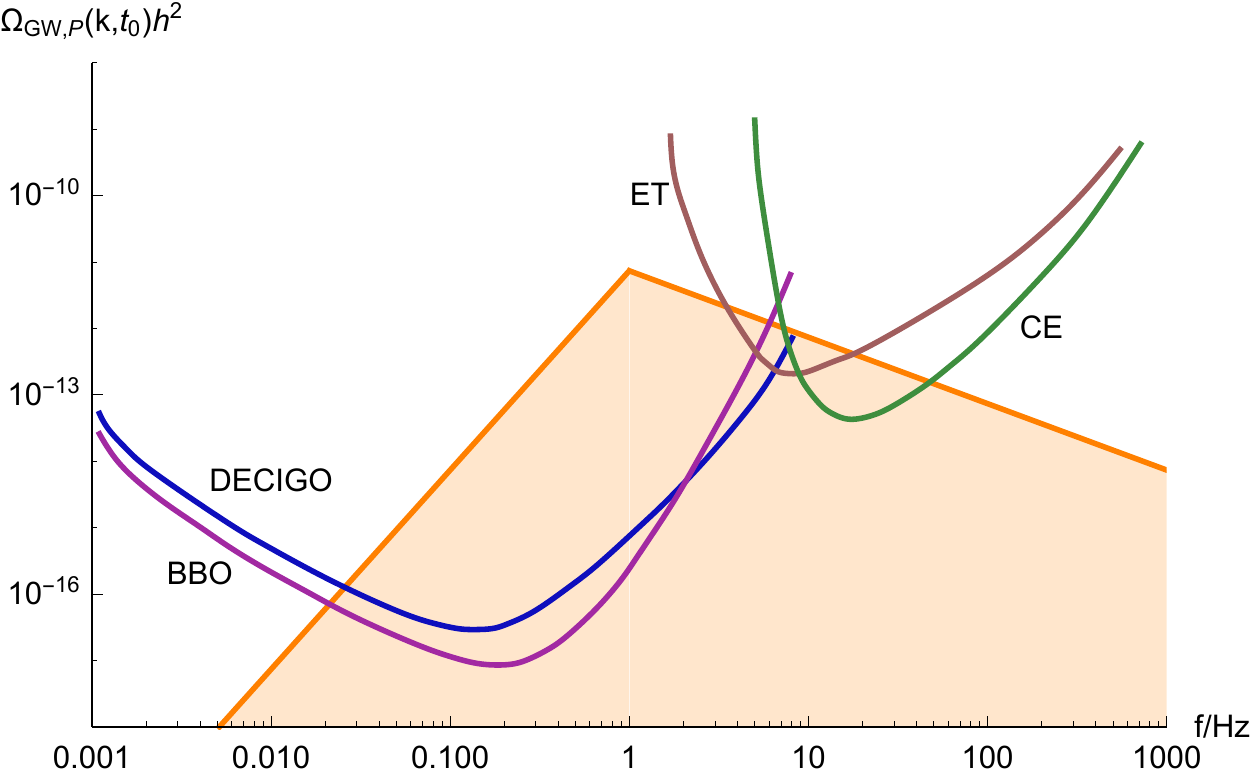}
	\includegraphics[width=3.2in]{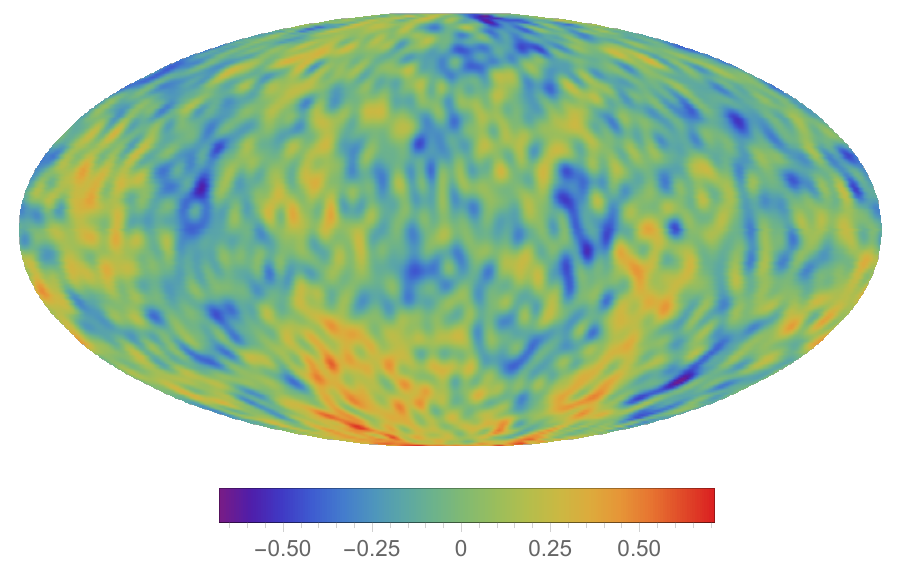}
	\caption{In the upper panel, the orange line presents $\Omega_{\mathrm{GW},P}(k,t_{0})h^{2}$ in the model discussed in Ref.~\cite{Takahashi:2008mu} with the parameters we choose, using the approximation method in Ref.~\cite{Saikawa:2017hiv}. This SGWB can be observed by DECIGO~\cite{Kawamura:2011zz}, BBO~\cite{Yagi:2011wg}, ET~\cite{Punturo:2010zz} and CE~\cite{Reitze:2019iox}.
	The lower panel shows the random realizations of the SGWB using the first 50 $l$-modes.}
	\label{fig:Omega}
\end{figure}
In Fig.~\ref{fig:Omega} we show the result of $\Omega_{\mathrm{GW},P}(k,t_{0})h^{2}$ and the random realizations of the SGWB. LISA~\cite{Audley:2017drz} and Taiji~\cite{Guo:2018npi} have the ability to detect such SGWB and its anisotropies. 

2) In the presence of monodromy, the discrete symmetry of the axion is explicitly broken by a quadratic term, and the effective potential reads $V(\phi)=\frac{1}{2}m^{2}\phi^{2}+\Lambda^{4}(1-\cos(\phi/\nu))$~\cite{Kaloper:2008fb,Hebecker:2016vbl,Silverstein:2008sg}. With the parameters $\nu=10^{-6}$, $\Lambda=3.6\times 10^{-18}$, $m=3\times 10^{-34}$, $\phi_{i}=H_{\mathrm{inf}}$, the tension of DWs is $\sigma=4.2\times 10^{-41}$, and the peak frequency and the GW energy spectrum peaks at $f=2.8\times10^{-9}\mathrm{Hz}$ with the peak value $\Omega_{\mathrm{GW},P}(t_{0})h^{2}=1.6\times 10^{-10}$, respectively. We find that the common-spectrum process observed by NANOGrav could be interpreted by the SGWB from DWs at the 68\% level. The SGWB produced in our model could be distinguished from the SGWBs from other sources~\cite{Ellis:2020ena,Blasi:2020mfx,Vaskonen:2020lbd,DeLuca:2020agl,Nakai:2020oit,Bian:2020bps,Addazi:2020zcj,Li:2020cjj} by characteristic large anisotropies with the improving sensitivity of pulsar-timing arrays.

\emph{Conclusion and discussion}.
In this letter, we have investigated the anisotropic SGWB from cosmic DWs when a discrete symmetry is spontaneously broken and the scalar field is light during inflation. Quantum fluctuations of the light scalar field remain constant at superhorizon scales, then induce large scale perturbations of energy density of cosmic DWs, finally lead to the anisotropic SGWB. The angular power spectrum in this scenario is larger than $10^{-2}$, which is a distinctive feature expected to be detected by the GW detectors. Since primordial GWs are too weak to be detected in low-scale inflationary models, observing the anisotropic SGWB provides a potential way to detect the inflationary energy scale even though it is several orders of magnitude lower than the grand unified theory scale.

Since $\rho_{\mathrm{DW}}$ at large scales is highly suppressed by $P(\tilde{\phi}<0, t)$, in principle $\Delta V$ is not required to prevent DWs becoming dominant. In this case, $\Omega_{\mathrm{GW},P}$ reaches the maximum when the increase of $\Omega_{\mathrm{GW}}$ and the decrease of $P(\tilde{\phi}<0, t)$ in Eq.~\eqref{eq:OmegaP} cancel each other out. However, the peak frequency is lower than $10^{-9}\mathrm{Hz}$, otherwise the GW signal is too weak to be observed. The anisotropic $k^{-1}$ slope of $\Omega_{\mathrm{GW},P}$ might be detected by SKA.

In the case of $H_{\mathrm{inf}}\gg \nu$, $\phi$ rolls down the potential and may cross the barrier at $\phi=0$ when DWs form, which implies our assumption $P(\tilde{\phi}<0, t)$ equals to the probability of one domain is violated. The linear approximation in  Eq.~\eqref{eq:deltaGW} will be replaced by a complicated form and non-Gaussianity arises. Even though the angular power spectrum is weakened and could be smaller than $10^{-2}$, it provides us a good chance to detect $H_{\mathrm{inf}}$ by the non-Gaussianity and relation between $l(l+1)C_{l}$ and $H_{\mathrm{inf}}/\nu$.

After the DW formation, the scalar field begins to oscillate around the minimum of the potential, leading to a resonant amplification of perturbations of $\phi$ inside the Hubble horizon, which is similar to the preheating scenario. Again relative to the difference between the averaged initial value in each large scale region, the amplified perturbations of $\phi$ will produce an extra anisotropic SGWB. The profile of $\Omega_{\mathrm{GW}}$ of such a SGWB contains useful information about $V(\phi)$, which helps us to further distinguish the particle physics model. For a larger $\nu$, for example string axions with $\nu\sim 10^{16}\mathrm{GeV}$, these GW signals are stronger and more likely to be detected.

The light field during inflation also contributes to entropy perturbations, depending on its decay products. In turn, we can use the detection of the SGWB to constrain entropy perturbations stronger than the CMB constraint.

In principle, our mechanism is applicable to cosmic strings.
Anisotropies of the SGWB from cosmic strings record abundant information of the source in a wild range of scale, 
detecting the frequency dependent angular power spectrum will help to reconstruct the potential and determine the inflationary model by detecting the slow time evolution of $H_{\mathrm{inf}}$. We left these interesting topics in future investigation.

\emph{Acknowledgments}
We thank Shaojiang Wang for fruitful discussions.
This work is supported in part by the National Natural Science Foundation of China Grants
No.11690021, No.11690022, No.11851302, No.11947302 No.11991052 and No.11821505,
in part by the Strategic Priority Research Program of the Chinese Academy of Sciences Grant No. XDB23030100 and by Key Research Program of Frontier Sciences, CAS.


\bibliography{anis_GW}

\end{document}